\documentclass[11pt,twocolumn,aps,prb,superscriptaddress,reprint,citeautoscript,sort&compress,square]{revtex4-1}

\usepackage{graphicx,amsmath,amssymb,color}
\usepackage{tabularx, ctable}
\usepackage{ulem}
\mathchardef\mhyphen="2D

\usepackage{numprint}

\newcommand{\SI}[1]{\textcolor{black}{#1}}

\usepackage[breaklinks=true,colorlinks=false,citecolor=blue,linkcolor=blue,urlcolor=blue] {hyperref}

\setcitestyle{super}

\begin{document}

\title{Single-photon detection using high-temperature superconductors}

\author{I. Charaev$^{*,+}$}
\affiliation{Massachusetts Institute of Technology, Cambridge, MA 02139, USA}
\affiliation{University of Zurich, Zurich 8057, Switzerland}
\author{D. A. Bandurin$^{*,+}$}
\affiliation{Massachusetts Institute of Technology, Cambridge, MA 02139, USA}
\affiliation{Department of Material Science and Engineering, National University of Singapore, 117575 Singapore}
\author{A. T. Bollinger}
\affiliation{Brookhaven National Laboratory, Upton NY 11973, USA}
\author{I. Y. Phinney}
\affiliation{Massachusetts Institute of Technology, Cambridge, MA 02139, USA}
\author{I. Drozdov}
\affiliation{Brookhaven National Laboratory, Upton NY 11973, USA}
\author{M. Colangelo}
\affiliation{Massachusetts Institute of Technology, Cambridge, MA 02139, USA}
\author{B. A. Butters}
\affiliation{Massachusetts Institute of Technology, Cambridge, MA 02139, USA}
\author{T. Taniguchi}
\affiliation{International Center for Materials Nanoarchitectonics, National Institute of Material Science, Tsukuba 305-0044, Japan}
\author{K. Watanabe}
\affiliation{Research Center for Functional Materials, National Institute of Material Science, Tsukuba 305-0044, Japan}
\author{X. He}
\affiliation{Brookhaven National Laboratory, Upton NY 11973, USA}
\affiliation{Department of Chemistry, Yale University, New Haven CT 06520, USA}
\author{I. Božović}
\affiliation{Brookhaven National Laboratory, Upton NY 11973, USA}
\affiliation{Department of Chemistry, Yale University, New Haven CT 06520, USA}
\author{P. Jarillo-Herrero}
\affiliation{Massachusetts Institute of Technology, Cambridge, MA 02139, USA}
\author{K. K. Berggren*}
\affiliation{Massachusetts Institute of Technology, Cambridge, MA 02139, USA}

\begin{abstract}
\textbf{The detection of individual quanta of light is important for quantum computation, fluorescence life-time imaging, single-molecule detection, remote sensing, correlation spectroscopy, and more. Thanks to their broadband operation, high detection efficiency, exceptional signal-to-noise ratio, and fast recovery times, superconducting nanowire single-photon detectors (SNSPDs) have become a critical component in these applications. The operation of SNSPDs based on conventional superconductors, which have a low critical temperature ($T_\mathrm{c}$), requires costly and bulky cryocoolers. This motivated exploration of other superconducting materials with higher $T_\mathrm{c}$ that would enable single-photon detection at elevated temperatures, yet this task has proven exceedingly difficult. Here we show that with proper processing, high-$T_\mathrm{c}$ cuprate superconductors can meet this challenge. We fabricated superconducting nanowires (SNWs) out of thin flakes of Bi$_2$Sr$_2$CaCu$_2$O$_{8+\delta}$ and La$_{1.55}$Sr$_{0.45}$CuO$_4$ /La$_2$CuO$_4$ (LSCO-LCO) bilayer films and demonstrated their single-photon response up to 25 and 8 K, respectively. The single-photon operation is revealed through the linear scaling of the photon count rate (\textit{PCR}) on the radiation power. Both of our cuprate-based SNSPDs exhibited single-photon sensitivity at the technologically-important  $1.5~\mu$m telecommunications wavelength. Our work expands the family of superconducting materials for SNSPD technology, opens the prospects of raising the temperature ceiling, and raises important questions about the underlying mechanisms of single-photon detection by unconventional superconductors.}
\begin{center}
$^{+}$ These authors contributed equally. \\
\end{center}

\end{abstract}

\maketitle
The detection of individual light quanta is critical for a broad range of applications in both quantum and classical realms. 
Quantum computation\cite{SPD4Qcomp1,SPD4Qcomp2,SPD4Qcomp3,SPD4Qcomp4,SPD4Qcomp5,SPD4Qcomp6}, metrology\cite{SPD4Metrology,bhargav2021metrology,giovannetti2011advances}, secure communication\cite{SPD4Qcomp7,SPD4Qcomm1,SPD4Qcomm2,SPD4Qcomm3}, single-photon imaging\cite{zhao2017single,xia2021short,ozana2021superconducting}, single-molecule detection\cite{SPD4MoleculeSens}, remote sensing\cite{SPD4RemoteSensing,zhu2017demonstration} and correlation spectroscopy\cite{carp2020diffuse,poon2022first} are just a short list of applications where faint light detectors play a crucial role. 
At present, the technology of single-photon detection is predominantly enabled by photo-multiplier tubes, single-photon avalanche diodes, visible light photon counters and superconducting detectors~\cite{ota2021photon,AvalancheReview,kim1999multiphoton,Berggren2013}.
In particular, superconducting-nanowire single-photon detectors (SNSPDs, also sometimes simply SSPDs) stand out in terms of broadband operation~\cite{wolff2021broadband} with high detection efficiency~\cite{reddy2020superconducting,Hu20,Chang}, exceptional signal-to-noise ratio~\cite{hochberg2019detecting}, ultra-high temporal resolution~\cite{korzh2020demonstration},  and fast recovery time\cite{cherednichenko2021low}. These exceptional characteristics are enabled by the physics behind SNSPD operation which is based on the generation of electrical signal (voltage pulse) triggered by the local suppression of the superconducting state upon photon absorption in a current-biased SNW~\cite{Engel_2015}. The formation of such a non-superconducting region (hot spot) leads to the local reduction of the maximum (critical) supercurrent. If the bias current, $I_\mathrm{bias}$, exceeds the critical value for this region, the SNW transitions to the resistive state, leading to the voltage spike~\cite{natarajan2012superconducting}. 


\begin{figure*}[ht!]
	\centering\includegraphics[width=\linewidth]{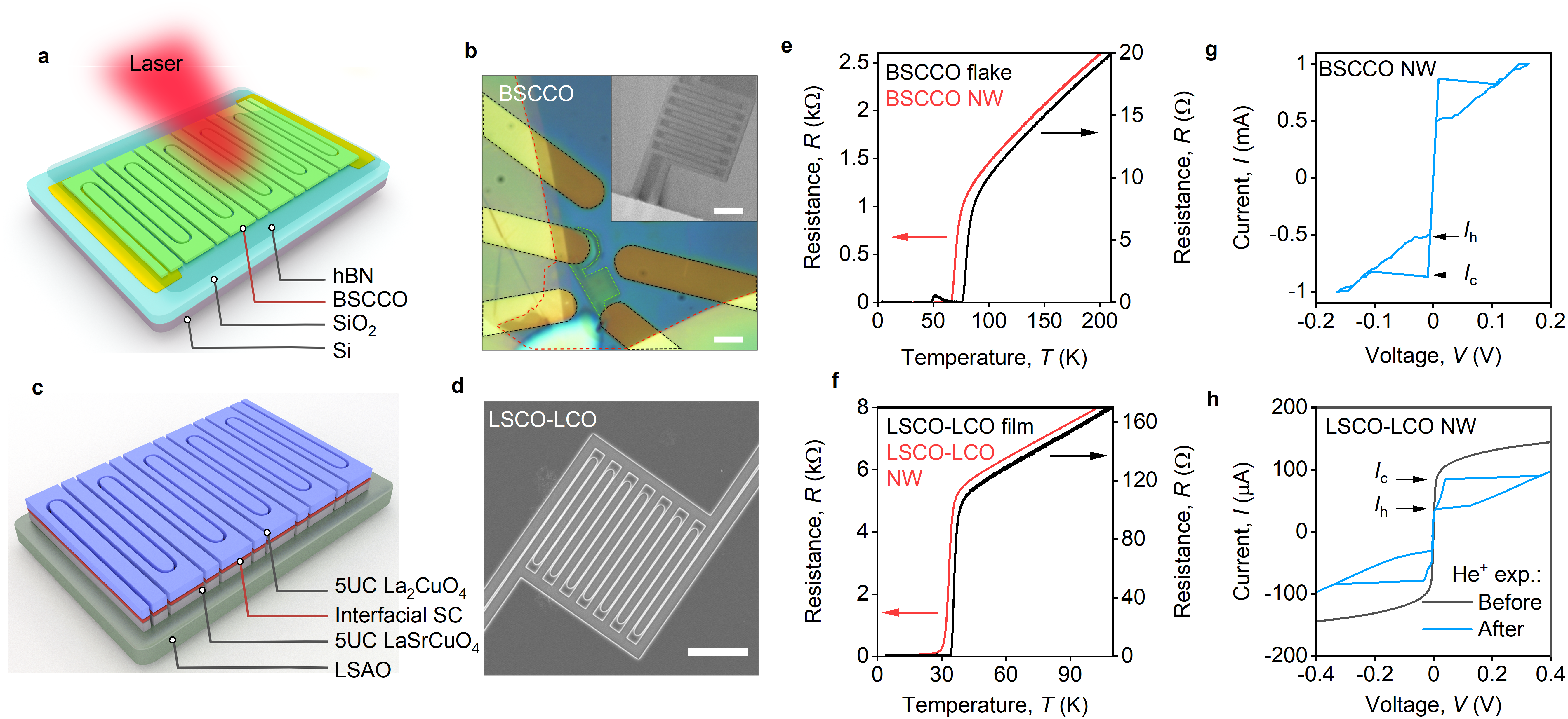}
	\caption{\textbf{High-$T_c$ superconducting nanowires.} \textbf{a,} Schematic of the BSCCO single-photon detector: A relatively thin flake of BSCCO is covered by a much thicker flake of hBN and transferred onto ultra-flat gold contacts. SNW region is defined by a helium beam exposure. \textbf{b,} Optical photograph of the BSCCO device. Scale bar is 3~$\mu$m.  Inset: Example of the SEM image of the BSCCO SNW produced by the He$^+$ beam exposure (similar but not identical to that from the photograph). The scale bar is 2~$\mu$m.  
	\textbf{c,} Schematic of the LSCO-LCO single-photon detector: High-$T_c$ two-dimensional superconductor is formed at the interface between the 5 UC of the LCO insulator and the 5 UC of the LSCO metal. 10 nm of chromium-gold was used for contact leads.  \textbf{d,} An SEM image of a typical LSCO-LCO SNW device. The scale bar is 2~$\mu$m. \textbf{e-f,} Examples of the $R(T)$ dependencies for BSCCO (e) and LSCO-LCO (f) flake, film and SNWs. \textbf{g,} Typical $I\mhyphen V$ curve for the BSCCO SNWs measured at $T=3.7~$K. \textbf{h,} Typical $I\mhyphen V$ curves of the LSCO-LCO SNWs measured at $T=3.7~$K before and after He$^+$ ion exposure.}
	\label{Fig_1}
\end{figure*}

The idea to use the nonlinearity of the superconductor-to-metal phase transition for detection dates back to sixties when the SNWs were proposed as a sensitive tool to detect elementary particles~\cite{sherman1962superconducting}. The evolution of this proposal led to the development of sensitive bolometers based on SNW made out of niobium nitride (NbN) thin films\cite{johnson1996bolometric} followed by a successful demonstration of the single-photon detection regime that gave birth to the SNSPD technology~\cite{semenov2001quantum, gol2001picosecond}.  At present, the SNSPDs require the use of costly and bulky cryocoolers to bring the SNWs below the liquid helium temperatures where the superconducting state of the SNW is well developed. For this reason, for the past two decades, there has been a continuous exploration of other superconductors with higher critical temperatures, $T_c$, that would allow for single-photon detection at elevated $T$~\cite{shibata2021review}. This search proved to be exceedingly challenging, and only one promising candidate emerged so far: in SNSPDs made of MgB$_2$ thin films with the $T_c=30~$K, the single-photon response in the optical range was demonstrated at $T\approx 10~K$\cite{velasco2016high}. Surprisingly, with few exceptions~\cite{andersson2020fabrication,ejrnaes2017observation,Lyatti2020}, cuprate superconductors remain largely unexplored in this context in spite of their record-high $T_\mathrm{c}$. The reason lies in the technological challenge to produce SNW devices out of thin cuprate films — most of them rapidly degrade upon conventional nanofabrication processing. Indeed, attempts to realize SNSPD on YBa$_2$Cu$_3$O$_{7 \mhyphen x}$ (YBCO) SNWs revealed no single-photon response even at liquid helium temperatures~\cite{Frenkel,andersson2020fabrication,ejrnaes2017observation}. 
In this work, we demonstrate SNSPDs fabricated out of two high-$T_c$ cuprate superconductors, Bi$_2$Sr$_2$CaCu$_2$O$_{8+\delta}$ (BSCCO) and La$_{2\mhyphen x}$Sr$_x$CuO$_4$ (LSCO). 

\textbf{Design and characterization of the cuprate SNWs.} To fabricate SNWs out of BSCCO, we mechanically exfoliated bulk BSCCO crystals to produce relatively thin flakes. To protect the flakes from the environment, the exfoliation was performed in the inert atmosphere of an argon-filled glovebox. The flakes were transferred onto pre-patterned ultra-flat gold contacts and covered with relatively thin slabs of hexagonal boron nitride (hBN) as illustrated in Fig.~\ref{Fig_1}a. We intentionally focused on 10-15 nm thick slabs (Fig.~\ref{Fig_1}a), since ultra-thin BSCCO flakes are prone to faster degradation~\cite{BisDecay1,BisDecay2,BisDecay3}. Thicker flakes also facilitate stronger absorption of incident light while maintaining quasi-2D superconductivity. Nevertheless, even few-unit-cells flakes are not amenable to standard nanofabrication processes, since their doping level changes upon contact with the environment, rendering the sample non-superconducting~\cite{BisDecay1,PK_BSCCO,BSCCO_Nature}. To circumvent this, we used He$^+$ ion beam to define the SNW patterns by introducing defects into the encapsulated BSCCO flake, causing the exposed regions to become insulating~\cite{cybart2015nano,HebeamYBCO,martinez2019superconducting,Gozar2017,seifert2021high} (Fig.~\ref{Fig_1}b). In contrast to patterning with heavier ions (e.g. gallium or xenon), the exposed regions were not etched under He$^+$ beam irradiation. The inset of Fig.~\ref{Fig_1}b shows a typical scanning electron microscopy (SEM) image of a partially-encapsulated BSCCO flake after the He$^+$ patterning: a meander-like SNW can be seen due to the enhanced contrast of the exposed regions. The total length of the BSCCO SNW was 56~µm whereas its width was of the order of 100~nm.

\begin{figure*}[ht!]
	\centering\includegraphics[width=1\linewidth]{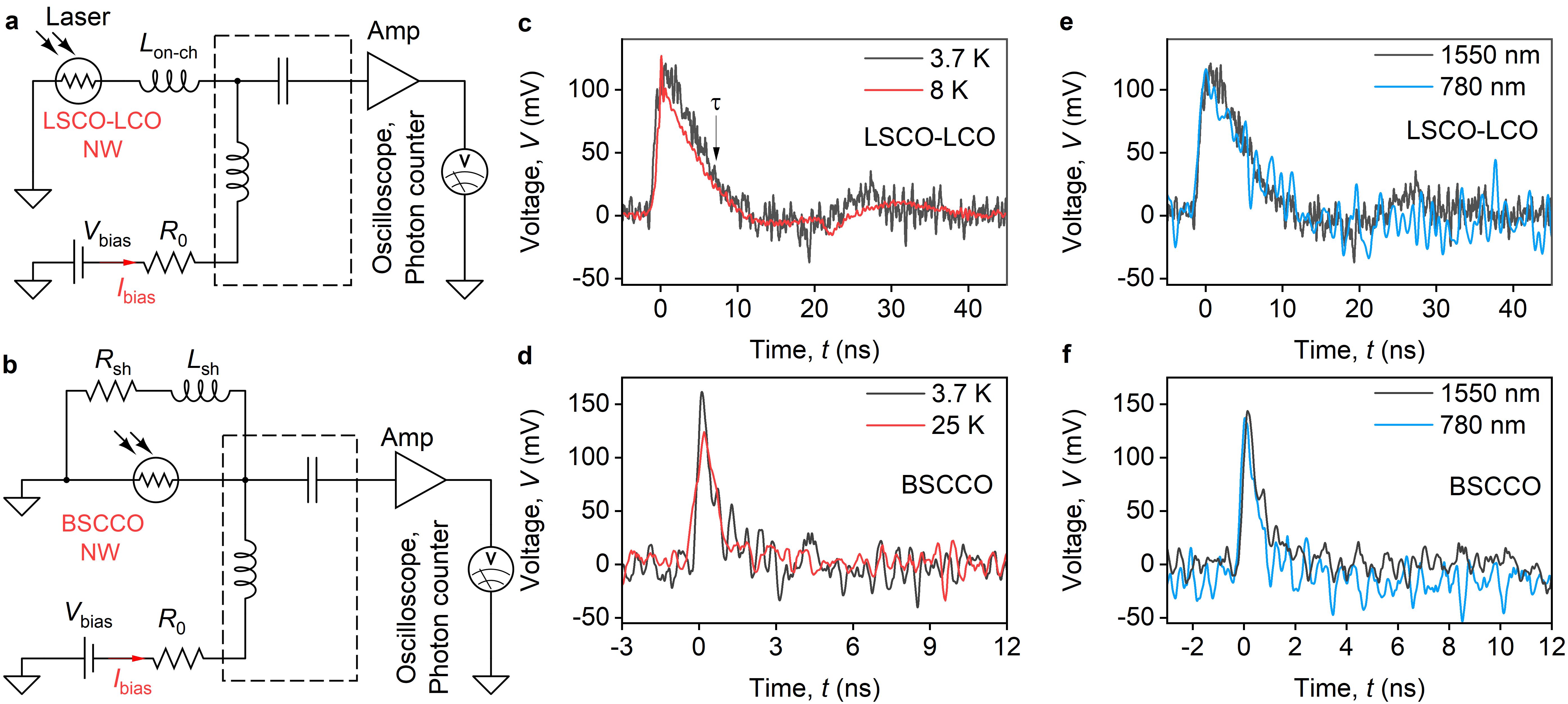}
	\caption{\textbf{Photovoltage generation in cuprate NW detectors.} \textbf{a,} The simplified circuit diagram used to measure the photoresponse of the LSCO-LCO SNW detector. The SNW is current-biased by an isolated voltage source connected to the DC port of the bias tee (dashed rectangle) through a resistor, $R_\mathrm{0}$. Incident radiation triggers a voltage spike generating a short pulse that propagates through the AC port of the bias tee to the preamplifier and is read out using an oscilloscope or a photon counter. $L_\mathrm{on-ch}$ is an on-chip kinetic inductor made out of the LSCO-LCO film. \textbf{b,} The simplified circuit diagram used to measure the photoresponse of the BSCCO SNW detector. $R_\mathrm{sh}$ and $L_\mathrm{sh}$ are the shunt resistor and the inductor connected in parallel with the BSCCO SNW to prevent it from latching.  \textbf{c-d,} Photovoltage $V_\mathrm{ph}$ pulses measured in the LSCO-LCO (c) and BSCCO (d) photodetectors at given $T$ and $\lambda=1.5~\mu$m. The devices are biased to the 95$\%$ of their critical current for given $T$. \textbf{e-f,} The $V_\mathrm{ph}$ pulses measured at given $\lambda$ for the LSCO-LCO (e) and BSCCO (f) devices at $T=3.7~$K and $T=16~$K respectively.} 
	\label{Fig2}
\end{figure*}

For successful implementation of the high-temperature SNSPD technology, it is critical to ensure large-scale production of thin high-$T_c$ films and their patterning using conventional nanofabrication processes. For these reasons, in this work we have also explored thin LSCO-LCO bilayer films. These heterostructures offer several advantages. First, high-quality LSCO-LCO bilayer films can be grown using atomic-layer-by-layer molecular beam epitaxy (ALL-MBE)~\cite{Gozar2008,LSCO-LCO_BL} (\SI{Supplementary Section 1}). Second, while in such bilayers neither constituent material is superconducting — LCO is an insulator and LSCO a non-superconducting metal — the heterostructure shows high-$T_c$ interface superconductivity confined to a single CuO$_2$ plane~\cite{Gozar2008,LSCO-LCO_BL}. Thus, using such heterostructures, high-$T_c$ SNWs can be fabricated with the cross-section reduced by two orders of magnitude, the smallest in any SNWs so far. Finally, LSCO-LCO bilayers are stable in air for years, and resilient to degradation during standard lithography, etching, and contact deposition processes. In this work, we used heterostructures comprised of 5-unit-cells(UC)-thick layer of LCO grown on top of 5-UC-thick layer of LSCO (Fig.~\ref{Fig_1}c). We used the electron-beam lithography to define an SNW meander structure typical for SNSPDs, 60 µm in length and 100 nm in width, with the filling factor 0.28 (Fig.~\ref{Fig_1}d). For the fabrication details, see Methods.

After fabrication, we characterized the transport properties of our high-$T_c$  SNWs. Figures~\ref{Fig_1}e-f show the typical temperature dependences of the resistance, $R(T)$,  of SNWs, fabricated out of BSCCO and LSCO-LCO, revealing their respective critical temperatures of 69.8 K and 34.4 K as determined from the maximum of the $dR/dT(T)$. The obtained values are somewhat lower than those obtained for the parent BSCCO flake and LSCO-LCO bilayer film, viz. 79.8 K and 35.5 K, respectively  (Figs.~\ref{Fig_1}e-f, black traces). This indicates a mild degradation of the materials’ superconducting properties during the fabrication. BSCCO underwent a more substantial change in $T_c$ likely because of its much stronger sensitivity to the environment. 

An important characteristic of most SNWs, enabling the generation of a voltage pulse upon single-photon absorption, is their metastable state that emerges under current biasing\cite{skocpol1974self}. This metastable state appears due to the competition between the current-induced Joule self-heating of the SNW in the normal state and electron cooling processes, and manifests itself as a pronounced hysteresis on the $I\mhyphen V$ characteristics. Figure~\ref{Fig_1}g shows a typical $I\mhyphen V$ curve measured in our BSCCO device below $T_c$ when the SNW is current-biased. A clear hysteretic behaviour characterized by the switching current $I_c=883~\mu$A and the retrapping current $I_h=516~\mu$A is observed. On the contrary, the $I\mhyphen V$ curves of the as-prepared LSCO-LCO SNWs were hysteresis-free and thus they were not suitable for a single-photon detector  \textit{per se}. To circumvent this problem, we exposed the device to a relatively small dose of He$^+$ ions~\cite{Gozar2017} ($10^{16}$ cm$^{-2}$). Such exposure had only a mild effect on $I_c$, $T_c$ and the normal state $R_s$; however it led to the desired $I\mhyphen V$ hysteresis with $I_c=86.7~\mu$A and $I_h=36.3~\mu$A. The origin of $I\mhyphen V$ hysteresis in the exposed LSCO-LCO bilayer SNWs is yet to be understood. 

\begin{figure*}[ht!]
	\centering\includegraphics[width=0.9\linewidth]{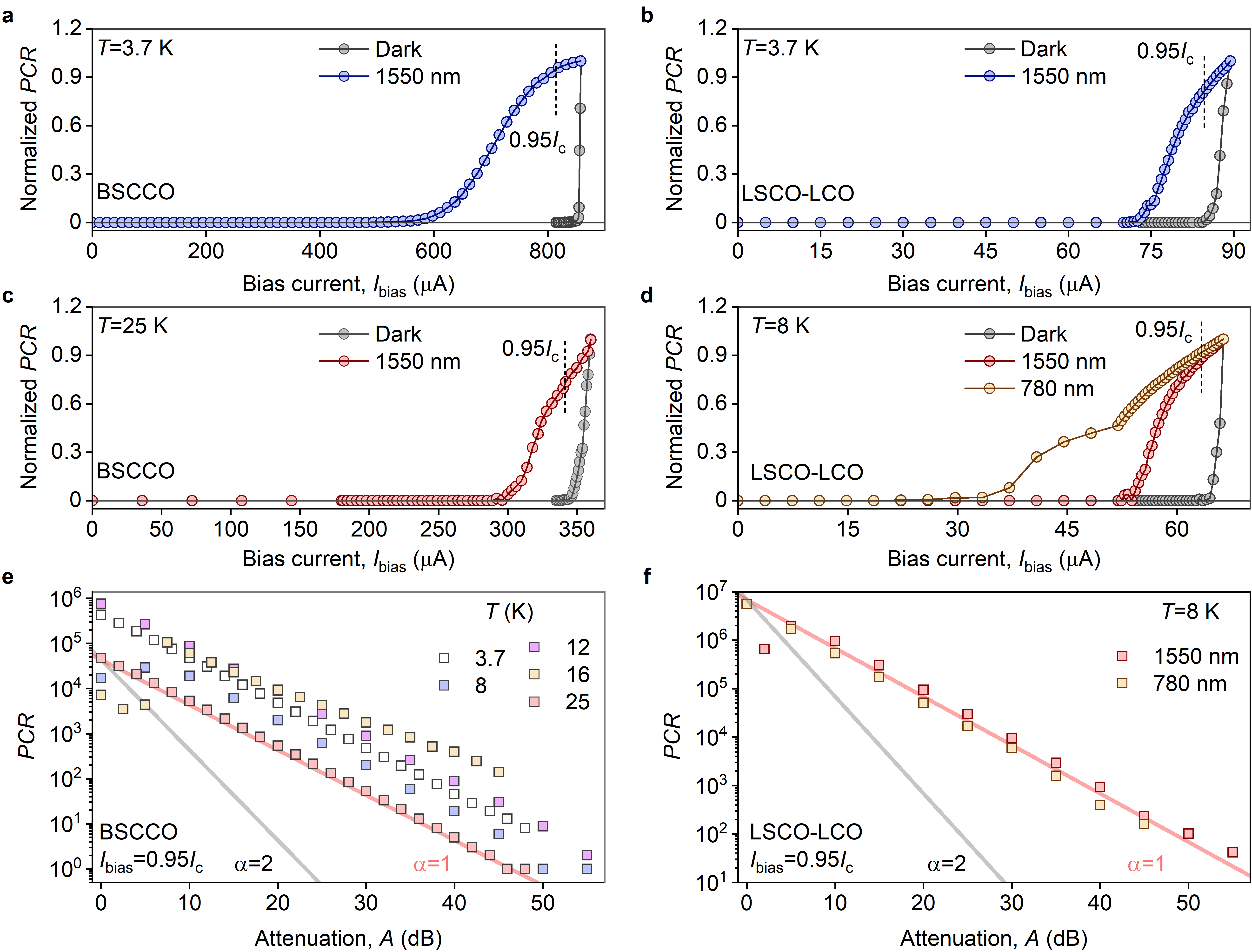}
	\caption{\textbf{Single-photon detection by cuprate SNWs.} \textbf{a-b,} Normalized to its maximum value \textit{PCR} as a function of bias current, $I_\mathrm{bias}$, measured in the BSCCO (a) and LSCO-LCO (b) photodetectors in the dark and upon illumination with $\lambda=1.5~\mu$m light. \textbf{c,} Same as (a) but for $T=25~$K. \textbf{d,} Normalized \textit{PCR} vs $I_\mathrm{bias}$ for different $\lambda$ measured in the LSCO-LCO device at $T=8$~K. Vertical dashed lines in (a-d) show $I_\mathrm{bias}$ that corresponds to the 95$\%$ of the switching current at $T=3.7~$K. \textbf{e,} \textit{PCR} vs attenuation factor, \textit{A}, at different $T$ for the BSCCO device at given $I_\mathrm{bias}$. $\lambda=1.5~\mu$m. Note, to avoid overheating of the BSCCO SNW at $T=25~K$ we intentionally reduced the initial output power of the laser, $P_0$, which resulted in the shift of the \textit{PCR}(\textit{A}) dependence towards smaller count rates. \textbf{f,} \textit{PCR} vs \textit{A} dependencies for the LSCO-LCO devices measured at given $\lambda$ and $I_\mathrm{bias}$. $T=8~$K. Red and black lines in (e) and (f): guides to the eye showing \textit{PCR}$\sim P^\alpha_\mathrm{inc}$ scaling for the single-photon ($\alpha=1)$ and multi-photon ($\alpha>1)$ regimes.
	}
	\label{Fig3}
\end{figure*}

\textbf{Photoresponse measurements.} To perform the photoresponse measurements, we mounted our cuprate SNWs in a variable-temperature cryostat equipped with RF coax cables and an optical fibers. The latter was held approximately 1 cm away from the device so that a defocused laser beam covered the whole device area. The simplified circuit diagrams, used for the photoresponse measurements, are shown in Figs.~\ref{Fig2}a-b. The LSCO-LCO device was measured in a conventional SPD configuration in which the SNW was biased through a DC input of the bias tee using an isolated voltage source connected in series with a resistor, $R_\mathrm{0}$. The AC output was connected to the low-noise amplifier whose output was fed to an oscilloscope or a photon counter (Figs.~\ref{Fig2}a). To mitigate latching effects in this superconducting photodetector, the SNW was connected in series with an on-chip kinetic inductor, $L_\mathrm{on-ch}$, also made out of the superconducting LSCO-LCO bilayer~\cite{zhao2017single,chiles2020superconducting}. The BSCCO measurement configuration was somewhat similar, but in this case, the latching effects were minimized using a more conventional approach, namely via shunting the SNW by an external resistor and an inductor connected in series with each other
(Fig.~\ref{Fig2}b). In Fig.~\ref{Fig_1}g-h we show the typical $R(T)$ dependencies and $I\mhyphen V$ curves measured without the shunt resistor/inductor in the BSCCO photodetector and without the on-chip inductor in the LSCO-LCO device.

Figure 2c shows an example of the photovoltage generation, $V_\mathrm{ph}$, measured across the current-biased LSCO-LCO SNWs when the device was exposed to the laser beam radiation of wavelength $\lambda=1.5~\mu$m. The $V_\mathrm{ph}(t)$ traces of this SNWs  shared common features with the photoresponse of conventional SNSPDs. After photon absorption, $V_\mathrm{ph}(t)$ spikes, and quickly reaches its maximum value. This is  followed by a much slower decay with the characteristic time $\tau$, often referred to as dead or recovery time, that depends on the total kinetic inductance, $L_\mathrm{k}$, of the superconducting circuit \cite{kerman2009electrothermal}. The measured value, $\tau\approx7~$ns (determined as the time when the signal dropped to the 30$\%$ of its maximum value) is in agreement with our measurements of the kinetic inductance in the LSCO-LCO bilayer films (See \SI{Supplementary Section 2}).  The $V_\mathrm{ph}$ spikes in the LSCO-LCO SNW device were observed below and above the liquid helium temperature and could be detected up to $T=8~$K. At higher $T$, LSCO-LCO SNWs did not exhibit an $I\mhyphen V$ hysteresis, and thus no voltage pulses were observed upon illuminating the SNW with low-intensity laser light (\SI{Supplementary Section 3}).

As compared to LSCCO-LCO, the $V_\mathrm{ph}$ pulses in biased BSCCO SNWs were observed up to much higher $T=25~$K (Fig.~\ref{Fig2}d) above which the $I\mhyphen V$ hysteresis disappeared (\SI{Supplementary Section 3}). The spikes  were characterized by a much faster recovery time $\tau\approx0.8~$ns. We attribute this short $\tau$ to a smaller total kinetic inductance of the BSCCO photodetector (\SI{Supplementary Section 2}). We have also tested the dependence of $V_\mathrm{ph}$ on the photon energy and found that both LSCO-LCO and BSCCO SNWs yielded spikes at both $\lambda=1.5~\mu$m and $\lambda=780~$nm (Figs.~\ref{Fig2}e-f). 

\textbf{Single-photon sensitivity of cuprate photodetectors.} To get further insight into the performance of our cuprate photodetectors, we recorded the photon count rate, \textit{PCR}, (the number of $V_\mathrm{ph}$ pulses per unit time) as a function of the bias current, $I_\mathrm{bias}$. Figure~\ref{Fig3}a shows the \textit{PCR} normalized to its maximum value, measured in our BSCCO  device in the dark and upon exposing it to the $\lambda=1.5~\mu$m laser light. In the dark, spontaneous voltage pulses emerge close to the critical current, a typical behaviour of SNSPDs. The absolute value of these dark counts did not exceed $10^3$ s$^{-1}$, comparable to the values in conventional NbN SNSPDs. Upon illumination, the counts appeared at much lower onset current $I_\mathrm{bias}=0.62~I_c$ whereas the \textit{PCR} showed some tendency to saturation upon approaching $I_c$. This saturation indicates high internal detector efficiency \cite{marsili2013detecting}. With increasing $T$ to 25~K, the onset current decreased together with $I_c$, as expected for conventional SNSPDs  (Fig.~\ref{Fig3}c). Furthermore, we found that the \textit{PCR} for a given $I_\mathrm{bias}$ was almost independent of $\lambda$ and featured similar \textit{PCR}($I_\mathrm{bias}$) functional dependencies for $\lambda=780~$nm and $\lambda=1.5~\mu$m (\SI{Supplementary Section 4}).

The \textit{PCR}-$I_\mathrm{bias}$ characteristics obtained for the LSCO-LCO photodetector at $\lambda=1.5~\mu$m were rather similar to those for BSCCO yet the counts emerged at much smaller $I_\mathrm{bias}$ due to a smaller critical current in this SNW (Fig.~\ref{Fig3}b,d). The onset current was about $0.75~I_c$ for both $T=3.7$ and 8~K. Notably, upon decreasing $\lambda$ to 780 nm, the functional form of the \textit{PCR}($I_\mathrm{bias}$) scaling changed drastically and featured a rather unusual dependence (Fig.~\ref{Fig3}d). First, the counts appeared at much lower $I_\mathrm{bias}\approx20~\mu$A. Then, the \textit{PCR} exhibit a tendency to saturation on increasing $I_\mathrm{bias}$ to $\approx52~\mu$A. Last, above this value, the \textit{PCR} started to ascend again suggesting two distinct operation modes of the LSCO-LCO photodetector. 

To reveal the single-photon regime of our cuprate photodetectors, we measured their \textit{PCR} as a function of the incident power, $P_\mathrm{inc}$, a standard test for single-photon-detector candidates. We used a calibrated optical attenuator and gradually decreased the photon flux incident on our cuprate photodetectors while recording the \textit{PCR} at different $T$ and $\lambda$.  Figure~\ref{Fig3}e shows the \textit{PCR} vs the attenuation factor, $A=10 \mathrm{log}(P_\mathrm{0}/P_\mathrm{inc})$, measured in the BSCCO device at the given $T$ and $I=0.95~I_c$ at which the dark counts are negligible (vertical dashed lines in Figs.~\ref{Fig3}a-d). Here $P_\mathrm{0}$ is the output laser power and $P_\mathrm{inc}$ is the laser power incident onto device. We observed a drop of the \textit{PCR} with decreasing $P_\mathrm{inc}$ that followed approximately the $PCR\sim P^\alpha_\mathrm{inc}$ scaling with $\alpha\approx1$ indicating a single-photon detection regime \cite{gol2001picosecond}. This linear dependence was observed for both BSCCO and LSCO-LCO devices with just a marginal deviation of $\alpha$ from unity at $T=16~$K in the BSCCO SNW (Fig.~\ref{Fig3}e), which we tentatively ascribe to an anomalously high background count rate in that run.

\textbf{Discussion and outlook.} While the data support the hypothesis of single-photon detection in these two materials, a number of features of the data are different from the observations in SNSPDs in conventional superconductors.  Specifically: (1) we observed unusual $I\mhyphen V$ characteristics in hysteretic LSCO-LCO SNWs (Fig.~\ref{Fig2}h); (2) we observed at one temperature (16 K) $\alpha < 1$ slope in the \textit{PCR}(\textit{A}) curve for BSSCO at high $A$; and (3) we observed unusual structure in the \textit{PCR}($I_\mathrm{bias}$) curves for the LSCO-LCO SNSPD.  Given the early stage of development of materials processing used in this work, such anomalies are not unexpected.  We expect significant non-uniformity in the patterning and processing, so different portions of the SNWs could be participating in the detection process at different $T$ and $I_\mathrm{bias}$.

Last, we consider the microscopic processes governing the single-photon response in cuprate SNWs. In general, the absorption of a photon by the SNW leads to the excitation of one electron into an empty state in the conduction band above the superconducting gap. This excited electron relaxes and disturbs the superconducting equilibrium state leading to the formation of a normal domain. While the hot spot dynamics in the current-biased SNWs is qualitatively understood, an accurate theoretical description of the hot spot evolution lacks completeness~\cite{semenov2021superconducting}. Nevertheless, all existing models rely on a set of parameters inherent to a parent superconductor: e.g. coherence length, penetration depth, electron-phonon coupling strength, diffusion coefficient, and specific heat capacities of electrons and phonons. In superconductors that proved to work in the SNSPDs minor variations show on these characteristics. However, the properties of unconventional cuprate superconductors are drastically different in both superconducting and normal phases~\cite{Keimer2015,RevModPhysLinR, Zhou2021}. First, cuprates are d-waves superconductors with nodes in the gap dispersion. Second, optimally-doped cuprates in their normal state are strange metals whose behaviour is governed by strong interactions~\cite{RevModPhysLinR}. In the previous models, the role of these properties have not been addressed. 

In this work, we demonstrated single-photon detection in high-$T_c$ cuprate SNWs at temperatures up to 25 K. This work invites further investigation of scaling of operation to higher temperature.  Additionally, it is surprising that these materials, which are very different from past examples, also exhibit single-photon detection, suggesting that the detection mechanism may need to be reconsidered. Finally, our work paves the way for future developments in the high$-T_\mathrm{c}$ superconducting devices and their integration into on-chip photonic quantum information circuits.

\section*{Methods}

\subsection*{Fabrication of BSCCO photodetectors}
To fabricate SNWs out of BSCCO we mechanically exfoliated bulk crystals and deposited cleaved flakes onto polydimethylsiloxane (PDMS) polymer stamps attached to the glass slide. To avoid the degradation of the flakes, the exfoliation was done in the inert atmosphere of the argon-filled glovebox. The flakes were then transferred onto pre-patterned on Si/SiO$_2$ substrates with  ultra-flat titanium/gold contacts and covered by relatively thick ($\sim 50~$nm) slabs of hexagonal boron nitride. 

The fabrication of ultra-flat contacts comprised several steps and relied on a lift-off-free procedure. First, thin layers of titanium (3 nm) and gold (25 nm) were evaporated onto the Si/SiO$_2$ wafer. Next, negative e-beam lithography was used to define a mask for selective removal of the metal outside the designated contact areas. This removal was done by a combination of Ar and O$_2$ etching. Resist residues were further removed by immersing the substrates into the N-methyl-pyrrolidone (NMP) solution and subsequent aggressive stripping using extensive O$_2$ plasma cleaning. We intentionally increased the typical dwell time of the plasma cleaning in order to reduce the thickness of the gold layer to 20 nm. 

To define superconducting SNWs out of prepared partially-encapsulated BSCCO devices by He$^+$ ion beam, we first ran the simulation of the ion collision damage using SRIM software. This allowed us to estimate characteristic doses needed to introduce a significant amount of defects into the BSCCO crystal lattice and suppress  superconductivity. Next, starting from the obtained values, we performed dose tests using Zeiss Orion Microscope with Raith pattern generator and determined the optimum parameters needed to define the desired SNWs.

\subsection*{Fabrication of LSCO-LCO photodetectors}
To fabricate SNWs out of LSCO-LCO bilayer we started from thin bilayer films grown by atomic-layer-by-layer molecular beam epitaxy (ALL-MBE) technique (See Supplementary Section 1). Titanium/gold contact pads were fabricated using standard e-beam lithography followed by metal deposition. In order to produce the SNWs out of LSCO-LCO bilayer films, another round of e-beam lithography was made, using high-resolution positive e-beam resist (ZEP520A). The patterns were then transferred onto the LSCO-LCO film by $Ar^+$ milling at a beam voltage of 300 V and Ar flow of 9 sccm for 10 minutes. The residual resist was removed by immersing the substrate in the warm NMP solution. 

After fabrication, we irradiated our SNWs with He$^+$ using Zeiss Orion Microscope equipped with a Raith pattern generator. The dose varied in the range from 10$^{15}$ to 10$^{20}$ ions/cm$^2$. The samples were then characterized using transport measurements in order to identify those with pronounced hysteresis in the $I\mhyphen V$ curves. For the photoresponse measurements, we chose the sample which featured the strongest hysteresis with the minimal suppression of $T_\mathrm{c}$.

\section*{Acknowledgments}
Work in the P.J.H. group was partly supported through AFOSR Grant FA9550-21-1-0319, through the NSF QII-TAQS program (Grant 1936263), and the Gordon and Betty Moore Foundation EPiQS Initiative through Grant GBMF9463 to P.J.H. D.A.B. acknowledges the support from MIT Pappalardo Fellowship. I.Y.P acknowledges support from the MIT undergraduate research opportunities program and the Johnson \& Johnson research scholars program. K.K.B. and group members acknowledge support from Brookhaven Science Associates, LLC award No. 030814-00001. K.W. and T.T. acknowledge support from JSPS KAKENHI (Grant Numbers 19H05790, 20H00354 and 21H05233). Thin film synthesis and characterization at Brookhaven National Laboratory was supported by the U.S. Department of Energy, Basic Energy Sciences, Materials Sciences and Engineering Division. H.X. was supported by the Gordon and Betty Moore Foundation’s EPiQS Initiative through Grant GBMF9074. We acknowledge valuable discussions with Sergio Rescia (BNL) and Gabriella Carini (BNL) and their significant help during the planning and development of this research work. The authors would like to thank J. Daley and M. Mondol of the MIT Nanostructures lab for the technical support related to electron-beam fabrication and helium ion microscopy. We also thank Frank Zhao (Harvard) and Owen Medeiros (MIT) for  helpful discussions.

\section*{Data availability}
The data reported in Figs. 1–3 can be found on Zenodo. The other data that support the findings of this study are available from the corresponding authors upon reasonable request.

\section*{Author contributions}
D.A.B. and I.C. conceived and designed the project. I.C. and D.A.B. performed the transport and photoresponse measurements. D.A.B., I.C. and I.Y.P. fabricated the devices. B.A.B., M.C. and I.C. designed the readout circuit. I.C. and D.A.B. analyzed the experimental data with the help from I.B., P.J.H. and K.K.B.. X.H., A.T.B. and I.B. synthesized and characterized LSCO-LCO bilayer films. T.T. and K.W. provided high-quality hBN crystals. D.A.B. and I.C. wrote the manuscript with input from all co-authors. P.J.H., I.B. and K.K.B. supervised the project. All authors contributed to discussions. 
\newline

\section*{Competing interests}
The authors declare no competing interests.
\newline

*Correspondence to: ilya.charaev@physik.uzh.ch, dab@nus.edu.sg, berggren@mit.edu, 
%
%
%

\end{document}